# Agile and Pro-Active Public Administration as a Collaborative Networked Organization

Wojciech Cellary

Department of Information Technology
Poznań University of Economics
ul. Mansfelda 4, 60-854 Poznań, Poland
+48 618480549

cellary@kti.ue.poznan.pl

Willy Picard

Department of Information Technology
Poznań University of Economics
ul. Mansfelda 4, 60-854 Poznań, Poland
+48 618480549

picard@kti.ue.poznan.pl

## ABSTRACT
In highly competitive, globalized economies and societies of always-on-line people intensively using the Internet and mobile phones, public administrations have to adapt to new challenges. Enterprises and citizens expect public administrations to be agile and pro-active to foster development. A way to achieve agility and pro-activity is application of a model of Collaborative Network Organizations in its two forms: Virtual Organizations (VO) and Virtual Organization Breeding Environments (VOBE). In the paper, advantages are shown of public administration playing a role of a Virtual Organization customer on the one hand, and a Virtual Organization member on the other hand. It is also shown how public administration playing a role of a Virtual Organization Breeding Environment may improve its agility and promote advanced technologies and management methods among local organizations. It is argued in the paper that public administration should provide a Virtual Organization Breeding Environment as a part of public services.

## Categories and Subject Descriptors
J.1 [**Computer Applications**]: Administrative Data Processing: *Business, Government.*

K.4.3 [**Computers and Society**]: Organizational Impacts: *reengineering, computer-supported collaborative work.*

## 1. INTRODUCTION
In the past two decades, the public administration has been an object of deep changes related with the implementation of e-governance. The main stimulus for these changes was the broad social acceptance of information technologies, especially the Internet and mobile phones. Public administration is evolving to take advantage of new possibilities following from information technologies. As a consequence, public administration has to a large extent to rethink and re-engineer the way it processes information, including formal documents, flowing among institutions of the public sector, as well as among public sector on the one side, and citizens and enterprises on the other. New possibilities of information flow has stimulated changes in the way organizations are interacting. Organizations are working more and more in a networked way, replacing the classical static linear supply chain of organizations by more complex and more dynamic graphs of relations among organizations. As a result, Collaborative Networked Organizations (CNO) emerged, i.e. organizations intensively collaborating via Internet with other organizations, to complement each other's skills and competences, to achieve the effect of synergy, and to increase added value produced. Two general types of CNOs are distinguished: Virtual Organizations (VO) and Virtual Organization Breeding Environments (VOBE).

The public administration has to follow the paradigm shift towards CNOs to align with business needs and to well serve citizens who intensively use the Internet. What is more, aiming at economic growth, the public administration should stimulate the creation of CNOs and to facilitate Small and Medium Enterprises to join CNOs.

The goal of this paper is to identify to which extent concepts related with CNOs and their adoption by the public administration may improve its agility and pro-activity.

In Section 2, the main concepts related with CNOs are defined and discussed. In Section 3, a possible relationships between public administration and virtual organizations, are discussed in the context of public administration agility. In Section 4, potential advantages following from the public administration playing the role of a Virtual Organization Breeding Environment are presented. Finally, Section 5 concludes the paper.

## 2. COLLABORATIVE NETWORKED ORGANIZATIONS
As a result of globalization, competiveness of most markets has drastically risen in the past decades. To survive, organizations have to focus on their core competences and to collaborate with other organizations to provide more complex services and products which are individualized and well adapted to various needs of local markets. To support collaboration of specialized organizations, new organizational structures have been proposed.

First, the concept of a *Virtual Enterprise* (VE) was invented to define a structure composed of various enterprises which, however, appeared as a single enterprise for its customers [1]. Then, the concept of a *Networked Virtual Organization* (NVO) has been proposed as an extension of a VE encompassing both business and administrative units. More precisely, a *Networked Virtual Organization* (NVO) is a set of business and/or administrative units mutually cooperating through the Internet, perceived on the market as if they were a single organization [2-4]. Next, the concept of an NVO has evolved to a *Virtual Organization* to encompass any type of organization, including Non-Governmental Organizations and any kind of associations. A *Virtual Organization* (VO) is defined as an operational structure consisting of different organizational entities, created for a specific business, administrative or social purpose, to address a particular opportunity [5].

Basing on the above concept of a VO, the concept of *Virtual Organization Breeding Environment* (VOBE, sometimes shortened to VBE in the literature) has been proposed as "an association of organizations and their related supporting institutions, adhering to a base long term cooperation agreement, and adoption of common operating principles and infrastructures, with the main goal of increasing their preparedness towards collaboration in potential virtual organizations" [6].

Organizations willing to cooperate with other organizations within a VO to address an opportunity have to:

o search and select right partners,
o standardize and integrate the IT infrastructure and applications among partners,
o define the principles ruling their cooperation,
o negotiate and sign a number of contracts related with the addressed opportunity.

A VOBE allows potential partners to pre-define elements of the above list before an opportunity occurs. Therefore, the creation of a VO is speeded up and simplified as only aspects specific to a given opportunity have to be defined during VO creation. As a consequence, VOBEs lead to more agile VO creation processes, with faster responses to opportunities.

As presented on path 1 of Figure 1, a VOBE enables partners to:

o pre-identify interesting partners of a potential VO from among the members of the VOBE,
o pre-define and agree on the principles ruling their potential cooperation within a VO,
o standardize IT infrastructure and applications common to all the VOs emerging from a given VOBE.

When an opportunity arises, a VO may be created in an easier way as organizations being members of the VOBE are already prepared for cooperation. The creation of a VO may be then reduced to:

o the selection of appropriate partners from among VOBE members,
o the negotiation of a contract based on pre-defined cooperation rules,
o the adaptation of IT infrastructure and applications basing on the ones common to all the VOBE members.

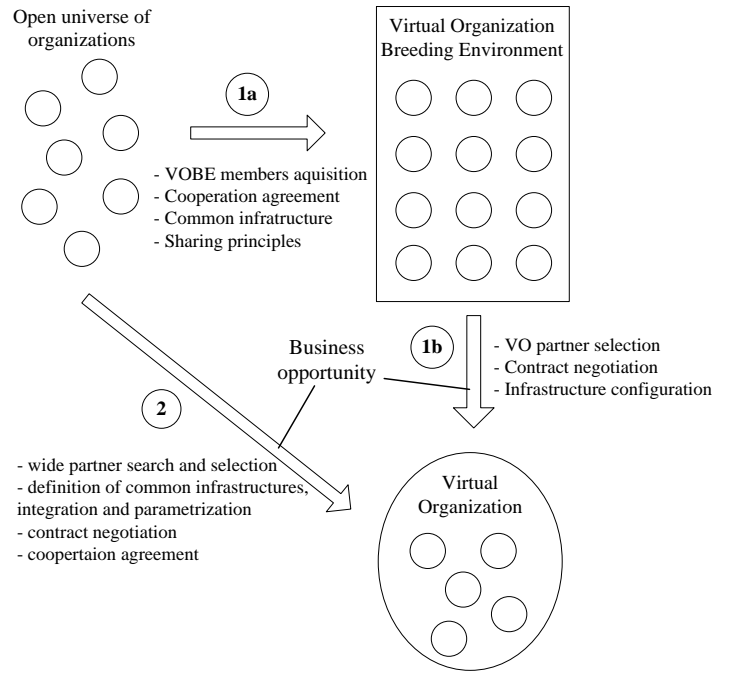

**Figure 1. Two approaches to the creation of virtual organizations (source ECOLEAD [7])**

The main aims of VOBEs are: establishment of mutual trust among organizations to facilitate their collaboration in VOs, reduction of cost and time necessary to find suitable partners for a particular VO, assistance in VO creation including reaching agreement between partners, and VO re-configuration aiming at adaptation to new challenges and opportunities.

While VOBEs are implemented in an ad hoc manner, *Service-Oriented Virtual Organization Breeding Environments* (SOVOBEs) are organized systematically around the concept of a service [8]. As a consequence, within the context of a SOVOBE, concepts underlying Service Oriented Architecture (SOA) may be applied at the coarser level of organizations. These concepts are the following ones:

o *service reuse* – a given organization may provide the same service within many VOs;
o *service abstraction* – the details of the implementation of services offered by a given organization within a VO are usually hidden for other organizations, because the implementation of the core business services is associated with the know-how capital that give the organization business advantage over competitors;
o *service discoverability* – services provided by organizations in a SOVOBE are described, so that both services and associated organizations may be identified as potential VO partners for a given opportunity;
o *service composition* – a complex service provided by a VO is a result of composition of services provided by VO partners and/or by the SOVOBE.

Services may be web services, potentially integrated by an Enterprise Service Bus (ESB), as well as services performed by humans. Depending on the type of service providers and consumers, the following classification of services is proposed (cf. Figure 2):

o *core member services* – services provided by SOVOBE members for chosen VO partners;

o *internal services* – services provided by the SOVOBE and consumed by its members. This set of services includes services for partner and service selection, tools for social protocol modeling, social network modeling, performance estimation, and competence modeling;

o *external services* – services provided by organizations operating outside the SOVOBE, but offered by the intermediation of the SOVOBE to its members. External services facilitate interactions between external organizations with SOVOBE, its members, and VOs;

o *façade services* – services provided by the SOVOBE to organizations outside the SOVOBE. Façade services provide external organizations with access to information about the SOVOBE and allow the submission of information to the SOVOBE. This set of services includes services for providing information about the SOVOBE and its members' profiles, and services for announcing market needs.

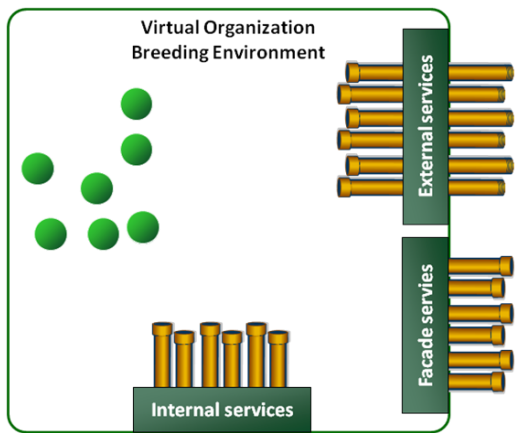

**Figure 2.** Services provided by a SOVOBE for its members and the outside world

To encompass the concepts of VE, NVO, VO, VOBE and SOVOBE, the term *Collaborative Network Organization* (CNO) has been coined [9]. A Collaborative Networked Organization is a network consisting of a variety of entities that are largely autonomous, geographically distributed, and heterogeneous in terms of their operating environment, culture, social capital and goals, which collaborate to better achieve common or compatible goals, and whose interactions are supported by computer networks.

## 3. AGILE PUBLIC ADMINISTRATION AND VIRTUAL ORGANIZATIONS

### 3.1 Public Administration as a VO Customer

Public administration is a major economical actor not only as a service provider, but also a service customer. Procurement of goods and services typically accounts for 10-15% of GDP for developed countries, and up to as much as 20% of GDP for developing countries [10].

A key element of each public tender is the Request for Tender (RFT). The product or service which is an object of the tender is specified in the RFT, together with the deadline for bid submission and the definition of evaluation criteria for submitted bids.

As an example, consider the city of Poznań, Poland that has organized a set of public tenders for the extension of the city stadium, related with the hosting of matches during the 2012 European Football Championship organized by the Union of European Football Associations UEFA, commonly referred to as Euro 2012 [11]. Each of these RFTs consists of the complete technical documentation of the construction works related with the extension of the stadium, the deadline for bid submission and criteria that have been used to evaluate submitted bids, and then to select the winner bid.

Additional requirements concerning potential bidders are frequently part of the RFT, reducing the set of organizations that may submit bids, such as requirements regarding former experience of organizations in the delivery of similar products or services. In the case of the Poznań tender for the extension of the first floor of the fourth stands of the city stadium [12], potential bidders are required to document their experience in stadium construction with attestation of at least one contract in the past five years concerning a construction site of $200 \text{ m}^2$ or more, and for a value greater or equal to 300,000 PLN (about 75,000 EURO).

Adding requirements concerning potential bidders to an RFT, e.g. properly documented former experience or recommendations from former clients, is a way to filter out organizations that are probably not able to deliver the product or the service specified in the RFT.

For so complex undertaking as the extension of a stadium, it is natural that bidders are – explicitly or implicitly – virtual organizations. Therefore, requirements concerning potential bidders should be extended to virtual organizations, e.g. requirements should be set on the number of VO members or the relationships among them. Such requirements may help public administration to filter out VOs that are probably not able to deliver the product or the service specified in the RFT due to inappropriate VO characteristics.

As an example, the city of Poznań could define requirements concerning the collaboration among potential VO members, e.g. at most 15 organizations should be involved in the VO responsible for the extension of the stadium, the structure of subcontractors should have at most 2 layers, etc.

### 3.2 Public Administration as a VO Member

In general, public administration is involved in two kinds of projects: repetitive and unique ones. With regards to the repetitive projects, public administration cumulate knowledge, so it is able to run them efficiently. Unique projects require various skills and competences that the public administration does not have and cannot acquire due to time constraints. The co-organization of the 2012 UEFA European Football Championship by Poland and Ukraine is an example of such unique project. The organization of this event requires, among others, the modernization of transportation and accommodation infrastructure, sport object extension or building from scratch, as well as the establishment of security and emergency plans. To face this unique challenge, the public administration should become a member of a VO encompassing also private companies and non-governmental organizations, to aggregate all necessary skills and competences. At a higher level, from the perspective of UEFA, the whole organization of the Euro 2012 may be considered as a single project performed by a VO composed of all necessary public, private and non-governmental organizations from Poland and Ukraine.

In general, public administration may cooperate in VOs not only with private organizations, but with both private and public units. As an example, consider the establishment of security plans concerning the Euro 2012 competition. On the one hand, it requires the cooperation of various public agencies at various levels of granularity, from the local to the international level. At the local level, the security plan requires involvement of hospitals, police and fire departments. At the international level it requires the involvement of border control and the management of hooligans coming from different countries. On the other hand, the establishment of security plans has to concern private companies too, for example bodyguard and security companies and private emergency medical companies.

Generally, the VO approach to public administration allows administration units to cooperate in a more agile way. The public administration may quite easily identify the core competences of its units, as well as private companies providing necessary complementary competences. The VO approach allows public administration to provide complex services by combining the core competences of its units on demand. As a consequence, the public administration is able to rapidly react to changes by "reconfiguring" itself", being therefore more agile.

## 4. PRO-ACTIVE PUBLIC ADMINISTRATION AS A SOVOBE

### 4.1 Public Administration as a Client-Centric Organization

A key element of successful e-governance implementation is the re-engineering of the public administration to center its functioning on the needs of its clients: citizens, enterprises and other organizations. Re-engineering consists of integration of services delivered by particular public administration units to reflect holistic processes realized by clients. Such processes are usually quite unique, so each of them needs a particular composition of services provided not only by public administration units, but also by private organizations, in general different organizations in case of each process.

If the VO paradigm is accepted by the public administration as its main way of providing services, then each compound service delivered by the public administration units and private organizations could be considered as a new VO. Then, the main goal of client-centric public administration should be defined as an ability to create VOs responsible for providing compound services corresponding to processes realized by its clients. Therefore, the public administration should become a SOVOBE.

As an example, the city of Poznań has to provide many complex services related with the organization of Euro 2012, among others, the extension of the stadium and the establishment of emergency plans. These services are answers to the needs of UEFA or other organizations, such as the Ministry of Interior and Administration. The delivery of each of these services requires the involvement of various public administration units and private organizations, in general different ones for each service. Therefore, it is reasonable for the city of Poznań to play the role of a SOVOBE supporting the breeding of VOs delivering services related with the organization of the Euro 2012 in Poznań, one compound service per VO. The same SOVOBE technology should be applied in all five Polish cites organizing matches of Euro 2012. After necessary adaptation following from different legal regulations, SOVOBE technology should be also applied in five Ukrainian cites responsible for Euro 2012 matches played in Ukraine. On a higher level, the Polish government should play the role of a SOVOBE supporting the breeding of VOs delivering higher-level services corresponding to the process of organization of Euro 2012 in Poland as a whole on demand of UEFA being the main client.

### 4.2 Public Administration as an Infrastructure Provider

Public administration has been historically the main infrastructure provider for the society: public administration has been responsible for the development and management of road networks, public transportation, water supply, gas supply, telecommunication networks, etc. The main reason for the public administration to be an infrastructure provider comes from the concept of public services: access to transportation means, to water, gas or phone has been considered as services that should be available to all the citizens. The public administration has been historically the warrant of the public character of these services and the underlying infrastructure.

Nowadays, by extension, the public administration should be a provider of the infrastructure associated with the creation of VOs delivering compound services, including but not limited to the ones provided by public administration units. As a consequence, the public administration should be a provider of the IT infrastructure including software shared by bred VOs.

Depending on the characteristics of the services delivered by the VOs, two main approaches to infrastructures are distinguished [13]:

- *cloud computing* – for the recurrent delivery of similar services,
- *service-oriented architecture* – for the reuse and aggregation of services in various, potentially unique contexts.

Cloud computing [14] is a business model of delivering IT resources and applications as services accessible remotely over the Internet rather than locally. In the traditional model, IT resources and applications are provided in the form of products which are sold or licensed from a vendor to a user and then exploited locally on a local IT infrastructure. According to the concept of cloud computing, instead of purchasing hardware or software, a user purchases remote access to them via the Internet. Cloud computing vendors invoice their customers on utility basis – pay as you go (such as electricity), or subscription basis (such as a newspaper).

With cloud computing, a uniform level of e-government practices may be achieved. From the point of view of the whole public administration, it is very important not to forget about the small and poor administration units, because the real benefits from electronic government can be achieved only when electronic public services are uniformly deployed throughout the whole state. The characteristics of the cloud computing model make it very well suited for organization of data processing in the whole public administration.

However, cloud computing is not well suited to the delivery of compound services whose parts are originating from different providers, which lack the repetitive character. In case of such services, the Service-Oriented Architecture [15] (cf. Section 2) is better adapted because it facilitates collaboration between independent administrative, business and other organizational units.

In the case of the Euro 2012, most services delivered during the organization of the event are rather unique, e.g. the stadium will be extended only once. On the contrary, after the event, many recurrent similar services will have to be delivered by various administrative units. As an example, the management of stadium by the public administration or, in many cases, by a private organization in a public-private partnership will be similar in Poznań, Warsaw, Gdańsk, and Wrocław in Poland as well as in Kiev, Donetsk, Lviv, and Kharkiv in Ukraine, cities in which Euro 2012 matches will take place. Therefore, while the organization of the Euro 2012 should be supported with a SOVOBE, with a strong focus on the service reuse and integration, the after-event management could take advantage of a cloud computing approach.

## 4.3 Public Administration as an Innovation Key Driver

Nowadays, in highly competitive economy, innovation is a key issue of sustainable growth. In every country enterprises are diversified with regard to innovations. Some enterprises are leaders of innovations, while others are lagging behind. For legal and organizational reasons, public administration cannot experiment with each innovation that appears on the market, so it is always less innovative than the business leaders. However, public administration should foster innovation among the conservative part of economy and society. The economic power of public administration as a major customer may be a lever to force citizens and private organizations to adopt innovative solutions, in particular new ways of communication and collaboration.

Public administration playing the role of a SOVOBE may foster innovation among enterprises, NGOs, and citizens by creating individualized VOs well adapted to local circumstances.

As an example, during the organization of the Euro 2012, the Polish governmental SOVOBE could enforce the use of multifunctional smart cards for all the services related with the Euro 2012. As a consequence, a single smart card could give and control access to transportation, stadiums, accommodation, telecommunication, museums, spectacles, and multitude of other services provided by Small and Medium Enterprises. Later on, all the public and private organizations that were supporting applications of multifunctional smart cards for the Euro 2012 would probably continue to do so. Moreover, the know-how gained by private and public organizations during Euro 2012 could be re-used in other sectors like for example the healthcare sector.

## 5. CONCLUSIONS

As follows from this paper, new forms of organization have arisen as a reaction to increased competition on globalized markets and to the new collaboration possibilities provided by information technologies [16]. Virtual organizations are taking advantage of modern collaboration forms to respond in an agile way to market needs, assembling the core competences of various organizations being members of a given VO. VO-related concepts do not have to be restricted to private organizations and may be applied to the public administration too leading to a more agile and proactive functioning of the public administration.

VO-based public administration is able to rapidly adapt its internal functioning to new challenges it faces by dynamically aggregating the competences of its administrative units in a unique manner. In most cases, agility of the VO-based public administration is not limited to the internal functions of the public administration, but it covers also collaboration with non-administrative organizations, namely private and non-governmental ones.

In this paper, it has been shown that re-engineering the public administration around the concepts related with CNOs leads not only to its greater agility, but also to its pro-active behavior. It happens when the public administration plays the role of a SOVOBE. Running a SOVOBE, the public administration not only provides services to the society, but also becomes a key driver of innovation.

Among future works, a model of federations of SOVOBEs, supporting the hierarchical structure of the public administration is still to be proposed. Another open issue concerns legal constraints imposed on involvement of public administration in CNOs in different countries.

## 6. ACKNOWLEDGMENTS

Our thanks to the Polish Ministry of Science and Higher Education for partial support of this research within the European Regional Development Fund, Grant No. POIG.01.03.01-00-008/08.